# FAULT TOLERANT MATRIX PENCIL METHOD FOR DIRECTION OF ARRIVAL ESTIMATION

Yerriswamy T.[1] and S.N. Jagadeesha[2]

[1]P.D. Institute of Technology, Visvesvaraya Technological University, Belgaum, Karnataka, India.
[2]J.N.N.College of Engineering, Visvesvaraya Technological University, Belgaum, Karnataka, India.
[1]swamy_ty@yahoo.com
[2]jagadeesha_2003@yahoo.co.in

## ABSTRACT

*Continuing to estimate the Direction-of-arrival (DOA) of the signals impinging on the antenna array, even when a few elements of the underlying Uniform Linear Antenna Array (ULA) fail to work will be of practical interest in RADAR, SONAR and Wireless Radio Communication Systems. This paper proposes a new technique to estimate the DOAs when a few elements are malfunctioning. The technique combines Singular Value Thresholding (SVT) based Matrix Completion (MC) procedure with the Direct Data Domain ($D^3$) based Matrix Pencil (MP) Method. When the element failure is observed, first, the MC is performed to recover the missing data from failed elements, and then the MP method is used to estimate the DOAs. We also, propose a very simple technique to detect the location of elements failed, which is required to perform MC procedure. We provide simulation studies to demonstrate the performance and usefulness of the proposed technique. The results indicate a better performance, of the proposed DOA estimation scheme under different antenna failure scenarios.*

## KEY WORDS

*Direction-of-Arrival, Matrix Completion, Matrix Pencil, Faulty Antenna Array, Singular Value Thresholding.*

## 1. INTRODUCTION

Antenna Array signal processing and estimation of Direction-of-Arrival (DOA) parameter of the received signal of interest, is of practical interest in RADAR, SONAR and wireless radio communication systems. There are many DOA estimation algorithms, these algorithms can be classified into three broad categories; Maximum Likelihood approach, subspace based approach and Direct Data Domain ($D^3$) approach [1] [2]. The Maximum likelihood (ML) method is optimal in the maximum likelihood sense [1], but requires accurate initializations to ensure global minimum and moreover, the method is highly computationally intensive. Expectation Maximization (EM) methods [3] [4] for ML approach reduces the computational requirements. However, the accurate initializations are still needed.

The subspace based algorithms like Multiple Signal Classification (MUSIC) [5] and Estimation of Signal Parameters via Rotational Invariance Technique (ESPRIT) [6], overcome the high computational requirement by exploiting the low rank structure of the noise free signal. These methods rely on the statistical properties of the data, and thus, need a sufficient large number of samples for accurate estimation. Furthermore, when the signal sources are highly correlated, the correlation matrix of the data tends to lose rank. This leads to the performance degradation of the DOA estimation algorithms based on the subspace approach. However, preprocessing scheme called spatial smoothing [7] is used to estimate the DOAs of the received highly correlated signal sources. Furthermore, there are recent algorithms derived for Nonuniform Antenna Arrays based on the above approaches [8] [9] and Ubiquitous Positioning [10].

Direct Data Domain ($D^3$) based algorithm namely, the Matrix Pencil (MP) Method [11] [12] is a practically attractive algorithm. The algorithm needs only one snapshot to estimate the DOAs and further, there is no need to form the correlation matrix. Correlated signal sources have no significant impact on the performance. The technique can also be applied to Nonuniform Antenna Array (NUA) without much modification [13]. However, the algorithm is sensitive to perturbation and measurement errors, resulting in a low Signal-to-Noise (SNR) threshold.

It is a common practice to use large number of elements in an antenna array, and hence, failure of a few elements will be disastrous in critical applications. Replacing the malfunctioning antenna array elements will be time consuming and costly. To deal with such problems, redundant antenna array is employed, which is wastage of the hardware and is a costly affair. The failure of elements results in incomplete data observations or sparse data observations. In such cases, conventional DOA estimation algorithms will have a problem to estimate the DOAs, because the structure and the statistical properties of the data cannot be found directly. If it is possible to reconstruct the complete data from the observed incomplete data, we can continue to estimate the DOA, from the ULA, where a few elements have failed to work, this will avoid redundant hardware.

For handling sensor failures, many modifications for conventional methods are proposed [14] [15]. Larson and Stoica [14], proposed a technique for estimating the correlation matrix of the incomplete data using the ML approach and has shown improvement in MUSIC for handling sensor failure. However, it increases the complexity. A method for DOA estimator to handle sensor failure based on neural network approach is proposed by Vigneshwaran et al [15]. The technique can handle correlated signal sources, avoids the Eigen decomposition. The drawback with these techniques is initialization of the network, and is performed by trial and error method. The authors in [16] proposed a DOA estimation technique by combining the EM algorithm with the MP method. The EM algorithm expects the missing data and maximizes the performance using the MP method. However, the method suffers from the drawback of increased complexity and requires good initialization.

Matrix Completion (MC) [17] is a process of completing the data matrix from incomplete data matrix. This problem arises in a variety of situations like system identification [18], collaborative filtering [19], DOA estimation [20], etc. The singular Value Threshold (SVT) algorithm proposed by Candes [21], can exactly recover the missing data from the knowledge of the location of the data elements in the matrix by solving a simple convex optimization problem [22], i.e., minimizing the nuclear norm, which is the sum of singular values of the data matrix.

This paper proposes a technique to deal with problem of DOA estimation when a few elements of the antenna array fail. The technique uses two procedures when the element failure is observed at time instance '$t$'. The first procedure, is to form the complete data matrix from the incomplete data matrix generated by the faulty ULA, using the SVT based MC algorithm. Later in the second procedure, DOAs are estimated directly from the obtained complete data using the MP method. The MC procedure first forms the location matrix. The location matrix describes the location of the elements that are functioning. The location matrix is formed by finding the distance between the samples at time instance '$t-1$' and at time instance '$t$'. At '$t-1$' all the elements are assumed to be working. The information from the location matrix, forms one of the input to the MC procedure using SVT.

The novelty in our algorithm is therefore present only when element malfunctioning is observed. The SNR and RMSE performance of the algorithm is evaluated in terms of varying percentage of element failures. The algorithm is compared with the standard MP Method. For the proposed technique, we consider the faulty ULA, whereas for the latter all the elements are assumed to be functioning.

The rest of the paper is organized as follows. The following section discusses the signal model for DOA estimation. In section 3, proposed technique is discussed followed by simulation results in section 4. Finally, conclusions are drawn in section 5.

## 2. SIGNAL MODEL

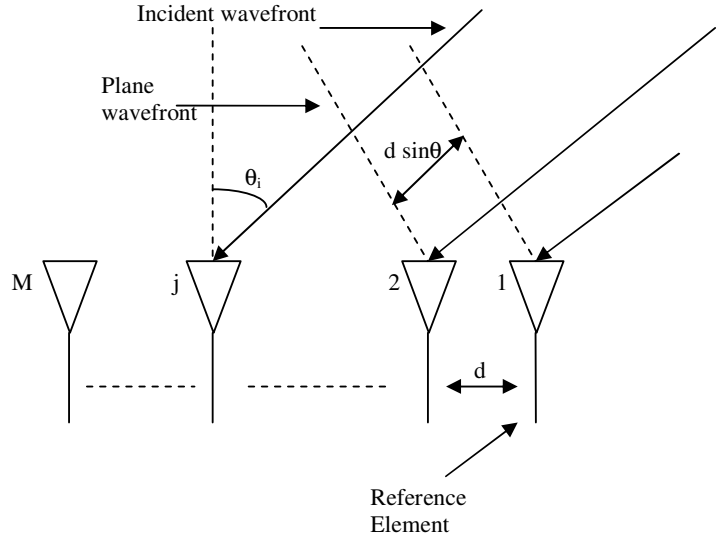

Figure 1. Uniformly spaced linear antenna array.

The DOA estimation problem is to estimate the directions of plane wave incident on the antenna array in presence of errors. The problem can be looked as parameter estimation. We here, mainly introduce the model of a DOA estimator. Consider a Uniform Linear Antenna Array (ULA) of '$M$' elements as shown in Figure 1. The elements are spaced equidistant '$d$' equal to '$\lambda/2$' from each other, where, '$\lambda$' is the wavelength of the signal received. The plane waves arrive at the array from direction '$\theta_i$' off the array broadside. The angle $\theta_i$ is called direction-of-arrival (DOA) of the received signal. Let $N$, narrowband signals $s_1(t), s_2(t), \cdots, s_N(t)$ impinge on the array with DOAs $\theta_i$, $i = 0,1,2, \cdots N-1$. Therefore, the signal $x_m$ received at the $m^{th}$ element at time instance '$t$' is

$$x_m(t) = \sum_{i=0}^{N-1} exp^{\left(-j\frac{2\pi}{\lambda}d(m-1)sin\theta_i\right)} s_i(t) + w_m(t) \qquad m = 1,2,\cdots,M \qquad (1)$$

Eq. (1) can be written in a compact form as

$$\mathbf{x}(t) = \mathbf{A}(\theta)\mathbf{s}(t) + \mathbf{w}(t) \quad t = 1,2,\cdots,T \qquad (2)$$

Where $T$ is the number of snapshots, considering only one snapshot i.e. $T=1$, $\mathbf{s}(t)$ is $M \times 1$ vector of signal sources, $\mathbf{w}(t)$ is $M \times 1$ noise matrix which is assumed to be Additive White Gaussian noise (AWGN), $\mathbf{A}(\theta)$ is the $M \times N$ array steering matrix and $\mathbf{x}(t)$ is the $M \times 1$ received signal vector.

Assume that at random locations a few elements are malfunctioning. In such case, the outputs from these elements are not available, resulting in the incomplete data vector $\mathbf{x}'(t)$ of size $M' \times 1$, where $M'$ is the number of elements functioning.

The objective in this paper is to continue the DOA estimation of $N$ source signals even when a few elements of a given ULA are malfunctioning.

## 3. PROPOSED DOA ESTIMATION ALGORITHM

The various steps of the proposed algorithm is described in Table 1 and illustrated in Figure 2. If there are element failures, we follow three important steps. First, we form the location matrix of the elements that are functioning. In second step, the information from the location matrix is used to form the complete data matrix using the SVT algorithm [21] (MC Procedure) and once the complete data is recovered, in the third step, the MP method is used to estimate the DOA. In the following section, the details of these steps are discussed.

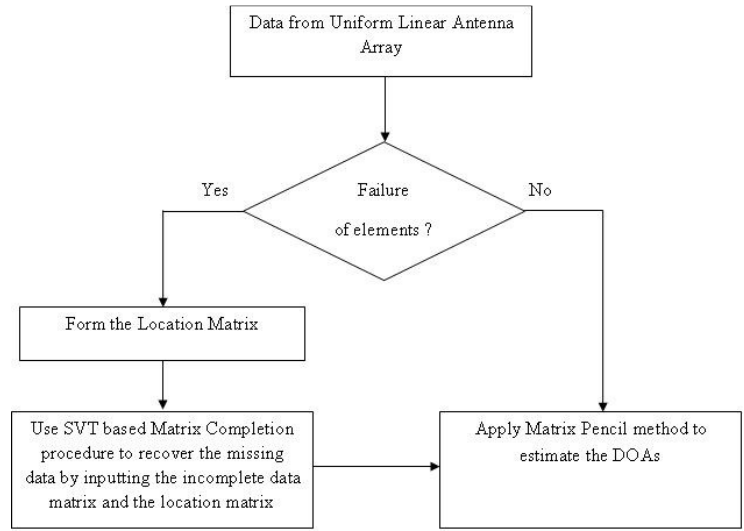

Figure 2. Basic idea of the proposed method

Table 1. Proposed Algorithm for DOA Estimation

| | |
|---|---|
| Assumption: <br> - All the $M$ elements are functioning at time instance $t-1$, the data from all the elements are collected <br> - Some number of elements are failed at time instance $t$ <br> - Number of signals impinging on the array $N$ <br> Begin | |
| Step1: | Collect the data snapshots at time interval t from all the elements |
| Step 2: | If length $(\mathbf{x}) \neq M$ then <br> Form the location matrix $\Omega$ <br> Apply SVT algorithm to complete the incomplete data matrix <br> Go to step 3. <br> Else |
| Step 3: | Apply MP method to estimate the DOA |

Signal & Image Processing : An International Journal (SIPIJ) Vol.2, No.3, September 2011

## 3.1 Array element location matrix

The matrix completion using the SVT algorithm requires the knowledge of the indices of the observed entries in the data matrix, which are the location of the functioning elements in the ULA. We propose a simple technique to locate the location of the working elements using previous sample and the present sample and comparing them to detect whether the particular element has failed or not. The algorithm to form a location matrix $\Omega$ is described in Table 2. and is based on finding the distance between the respective samples.

Table 2. Algorithm for forming Location Matrix

| | |
|---|---|
| Initialization : Index j = 0 Begin: | |
| Step1: | Collect and store snapshot $x_i(t-1)$, $i$ is the location value of the element |
| Step 2: | Collect and store snapshot $x_i(t)$ |
| Step 3: | Calculate $dist_i(x_i(t-1), x_i(t)) = |x_i(t-1) - x_i(t)|$ <br> If $dist_i(x_i(t-1), x_i(t)) < \epsilon$ \\ where $\epsilon$ is some threshold = 10$^{-2}$ <br> then <br> $\Omega_j = i$; \\ Element $i$ is functioning <br> else <br> $\Omega_j = i$; \\ Element $i$ is malfunctioning <br> End if; <br> j=j+1 |

## 3.2 Algorithm for DOA estimation

At time instance $t$, a few of the elements will fail to work, resulting in incomplete data. To recover the complete data from the observed incomplete data which resulted from failure of a few elements, we apply the SVT matrix completion algorithm given in Table 3.

Recovering a matrix from a sample of its entries is known as the matrix completion problem. In [16] [20], Candes and Recht, proved that the most of the low rank '$r$' matrices can be recovered from its partial set by solving a simple convex optimization problem.

$$\min_{\mathbf{x}'=P_\Omega(\mathbf{x})} \|\mathbf{x}\|_* \tag{3}$$

Where, $\|\mathbf{x}\|_*$ is called the nuclear norm and is defined as the sum of its singular values, **x** is the matrix to recovered, $\Omega$ is the set of indices's of the sampled entries, $P_\Omega(.)$ is a masking operator which selects the entries of **x** that are within $\Omega$ and $\mathbf{x}'$ is the collected partial samples. Provided that the number of samples obey $M \geq CM^{1.2}r\log M$ for some positive numerical constant $C$. The SVT algorithm developed by Cai, Candes and Shen in [21] is used to solve the norm minimization problem in (3). However, the entries must be selected randomly and we cannot hope to complete the matrix if some of the singular vectors of the matrix are extremely sparse.

If the singular vectors of $\mathbf{x}'$ are sufficiently spread, then there is a unique low rank matrix which is consistent with the observed entries. In such cases, one could, in principle, recover the unknown matrix by solving

$$\min_{\mathbf{x}'=P_\Omega(\mathbf{x})} \text{rank}(\mathbf{x}) \tag{4}$$

Unfortunately this is NP-hard. A popular alternative to the NP hard problem is the convex relation [17] given in (3) where, nuclear norm is the tightest convex relation of the NP-hard rank minimization problem.

The inputs to the SVT algorithm are, a parameter $\tau$, step size $\delta$, sampled set $\Omega$, sampled entries $\mathbf{x}(\Omega)$ and initializing vector $\mathbf{Y}^0 = 0$. The algorithm is

$$\mathbf{x}^k = shrink(\mathbf{Y}^k, \tau) \tag{5}$$
$$\mathbf{Y}^k = \mathbf{Y}^{k-1} + \delta_k \Omega^{\mathrm{T}}(\mathbf{x}' - \mathbf{x}^k)$$

Repeat steps in (5) until convergence. *shrink(.)* is a nonlinear function which applies soft thresholding rule at level $\tau$, to the singular values of the input matrix. The key property here is that for large values of $\tau$, the sequence $\{\mathbf{x}^k\}$ converges to a solution which very nearly minimizes (3). Hence, at each step, one needs to compute only atmost one singular value decomposition and perform a few elementary matrix additions.

For shrinkage operator, consider the SVD of **x** of rank r,
$$\mathbf{x} = \mathbf{U}\mathbf{S}\mathbf{V}^H, \quad \mathbf{S} = diag(\{\sigma_i\}, 1 < i < r) \tag{6}$$

Where **U** and **V** are the right and the left singular vectors and **S** is the singular value matrix. For each $\tau \geq 0$, a soft thresholding operator $D_\tau$ defined as follows

$$D_\tau(\mathbf{x}) = \mathbf{U}D_\tau(\mathbf{S})\mathbf{V}^H, \quad D_\tau(\mathbf{S}) = diag(\{(\sigma_i - \tau)_+\}, 1 < i < r) \tag{7}$$

Where $(*)_+$, is the positive part of the *. In other words, this operator simply applies the soft thresholding rule to the singular values of **x**, effectively shrinking these towards zero.

The problem of estimating the DOA from incomplete observations in our case can be written as

$$\min_{\mathbf{x}' = \mathbf{x}(\Omega)} \|\mathbf{x}\|_* \tag{8}$$

$\mathbf{x}'$, is the data collected from the working elements and $\Omega$ is the location matrix whose entries are the location of the elements that are working in a ULA. The operator $\mathbf{x}(\Omega)$ represents the data collected from only the working elements. Once the complete data is recovered, MP method is followed to estimate the DOAs [11] [12]. The SVT algorithm for recovering the missing data is shown in Table 3.

Table 3: Algorithm of SVT

| Input: | sampled set or location vetor $\Omega$, incomplete data x', step size $\delta$, tolerence $\epsilon$, parameter $\tau$, $k_{max}$ maximum iterations. |
|---|---|
| Output: | Recovered data **x** |
| | For k = 1 to $k_{max}$<br>  Let $s_k = \{\sigma_1^k, \cdots \sigma_s^k\}$ are s singular values and $r_k = rank(\mathbf{x}^k)$<br>  Set $s_k = r_k + l$<br>  Repeat<br>    Compute $[\mathbf{U}^{k-1}, \mathbf{S}^{k-1}, \mathbf{V}^{k-1}]_{s_k}$<br>    Set $s_k = s_k + l$<br>  until $\sigma_{s_k-l}^{k-1} \leq \tau$<br>  Set $r_k = \max\{j: \sigma_j^{k-1} > \tau\}$<br>  Set $\mathbf{x}^k = \sum_j^{r_k}(\sigma_j^{k-1} - \tau)\mathbf{u}_j^{k-1}\mathbf{v}_j^{k-1}$<br>if<br>    $\|\Omega^{\mathrm{T}}(\mathbf{x}' - \mathbf{x}^k)\|_F / \|\mathbf{x}'\|_F \leq \epsilon$ then break<br>set<br>    $\mathbf{Y}^k = \begin{cases} 0 & \text{if } (i,j) \notin \Omega \\ \mathbf{Y}^{k-1} + \delta(\mathbf{x}' - \mathbf{x}) & \text{if } (i,j) \in \Omega \end{cases}$<br>end for k<br>    $\mathbf{x} = \mathbf{x}^k$ |

Signal & Image Processing : An International Journal (SIPIJ) Vol.2, No.3, September 2011

For MP method, the signal model given in (1) is rewritten as,

$$x_m = \sum_{n=0}^{N-1} s_n z_n^{m-1} + w_n \qquad m = 1,2,\cdots,M \tag{9}$$

Where,

$$z_m = exp\left(j\frac{2\pi}{\lambda} d \sin \theta_n\right) \tag{10}$$

For simplicity the index *t* is eliminated. The MP method for noiseless data the algorithm begins by choosing a parameter $L$, known as pencil parameter. A good choice of this parameter is in between M/3 and 2M/3 [12].

Now we construct a matrix **X** from the data samples of (9)

$$\mathbf{X} = \begin{bmatrix} x_1 & x_2 & \cdots & x_{L+1} \\ x_2 & x_3 & \cdots & x_{L+2} \\ \vdots & \vdots & \ddots & \vdots \\ x_{M-L} & x_{M-L+1} & \cdots & x_M \end{bmatrix} \tag{11}$$

The two matrices $\mathbf{X_1}$ and $\mathbf{X_2}$ are defined as

$$\mathbf{X_1} = \begin{bmatrix} x_1 & x_2 & \cdots & x_L \\ x_2 & x_3 & \cdots & x_{L+1} \\ \vdots & \vdots & \ddots & \vdots \\ x_{M-L} & x_{M-L+1} & \cdots & x_{M-1} \end{bmatrix} \tag{12}$$

$$\mathbf{X_2} = \begin{bmatrix} x_2 & x_3 & \cdots & x_{L+1} \\ x_3 & x_4 & \cdots & x_{L+2} \\ \vdots & \vdots & \ddots & \vdots \\ x_{M-L+1} & x_{M-L+2} & \cdots & x_M \end{bmatrix} \tag{13}$$

The matrix pencil for two matrices $\mathbf{X_1}$ and $\mathbf{X_2}$ are defined as the linear combination of the two matrices with scalar β described by

$$\mathbf{X_1} - \beta \mathbf{X_2} \tag{14}$$

In the absence of noise, it is easy to verify that $\mathbf{X_1}$ and $\mathbf{X_2}$ can be decomposed in to $\mathbf{X_1} = \mathbf{AZB}$ and $\mathbf{X_2} = \mathbf{AB}$ such that

$$\mathbf{A} = \begin{bmatrix} 1 & 1 & \cdots & 1 \\ z_0 & z_1 & \cdots & z_{N-1} \\ \vdots & \vdots & \ddots & \vdots \\ z_0^{M-L-1} & z_1^{M-L-1} & \cdots & z_{N-1}^{M-L-1} \end{bmatrix} \times \begin{bmatrix} s_0 & 0 & \cdots & 0 \\ 0 & s_1 & \cdots & 0 \\ \vdots & \vdots & \ddots & \vdots \\ 0 & 0 & \cdots & s_{N-1} \end{bmatrix} \tag{15}$$

$$\mathbf{B} = \begin{bmatrix} z_0^{L-1} & z_0^{L-2} & \cdots & 1 \\ z_1^{L-1} & z_1^{L-2} & \cdots & 1 \\ \vdots & \vdots & \ddots & \vdots \\ z_{N-1}^{L-1} & z_{N-1}^{L-2} & \cdots & 1 \end{bmatrix} \tag{16}$$

$$\mathbf{Z} = \begin{bmatrix} z_0 & 0 & \cdots & 0 \\ 0 & z_1 & \cdots & 0 \\ \vdots & \vdots & \ddots & \vdots \\ 0 & 0 & \cdots & z_N \end{bmatrix} \tag{17}$$

provided that

61$$N \leq L \leq M - L, \quad M \text{ even} \qquad (18)$$

$$N \leq L \leq M - L + 1, \quad M \text{ odd}$$

the pencil is of rank $N$. Under this condition each value of $\beta = z_n$ is a rank reducing number of the pencil. However if $L$ is not within the above range, then none of $z_n$'s are a rank reducing number of a matrix pencil. This implies that the values of $z_n$'s are the generalized eigenvalues of the matrix pair $[\mathbf{X_1 X_2}]$. Furthermore, it can be shown that the generalized eigenvalues of $\mathbf{X_1} - \beta\mathbf{X_2}$ can be found from the non-zero eigenvalue of $\mathbf{X_2^\dagger X_1}$ where † is the Moore-Penrose pseudoinverse. Finally the DOAs are estimated by using equation.

$$\theta_n = -arg\sin\left(\frac{c}{d} arg(z_n)\right) \qquad (19)$$

Where, $c$ is the propagation velocity. In the absence of noise, the pencil will have rank $N$, which is not satisfied when the signal is corrupted by noise. To mitigate the effect of noise Total Least Squares (TLS) [23] approach is applied by taking Singular Value Decomposition (SVD) of $\mathbf{X}$ in (11). This method is known as Total Least Squares – Matrix pencil (TLS-MP) method.

We start with taking the SVD of $\mathbf{X}$ in (11). Only $N$ of singular values of this matrix corresponds to signals, while the rest corresponds to the noise. We generate a new filtered version of the data matrix.

$$[\mathbf{\acute{X}}] = [\mathbf{\acute{U}}][\mathbf{\acute{\Sigma}}][\mathbf{\acute{V}}] \qquad (20)$$

Where, $[\mathbf{\acute{U}}]$ are the first $N$ left singular vectors, $[\mathbf{\acute{V}}]$ are the first $N$ right singular vectors and $[\mathbf{\acute{\Sigma}}]$ are the first $N$ singular values. Now $\mathbf{X}$ is replaced by $\mathbf{\acute{X}}$ in (11). The above steps are followed to estimate the DOA of the received signals.

## 4. SIMULATION RESULTS

In this section, we examine the performance of the proposed DOA estimation technique using several simulations and compare with the standard TLS-MP method for various noisy conditions and elements failure scenario. The simulation is carried on Intel Core 2 Duo, 2.8 GHz processor, 2 Mb RAM, running on Windows XP SP2 16 bit operating system. We consider a ULA of 100 elements, the spacing between the elements is $\lambda/2$, where, $\lambda$ is the wavelength of the signal of interest. The signals are assumed to be impinging from the direction $\theta_i$ off the broadside of the array.

The signals are assumed to be complex exponential sequence given by

$$s(t) = exp^{(j\phi(t))} \qquad (21)$$

Where $\phi$ is the random phase uniformly varying between $[-\pi, \pi]$. The SNR in dB at each element is defined as

$$SNR = 10 \log_{10}(\sigma_s/\sigma_w) \qquad (22)$$

Where $\sigma_s$, is the signal power and $\sigma_w$, is the noise power. The Root Mean Square Error is defined as

$$RMSE = \sqrt{E\left[(\theta_i - \hat{\theta}_i)^2\right]} \qquad (23)$$

Where, $\theta_i$ is actual DOA and $\hat{\theta}_i$ is the estimated values. The failure of elements is considered to be at random locations. Six signals of equal magnitude are assumed to be impinging on the array from the directions $\theta_i = \{0°, 5°, 10°, 15°, 20°, 30°\}$.

Signal & Image Processing : An International Journal (SIPIJ) Vol.2, No.3, September 2011

In our first example, we compare the TLS-MP method and the proposed modified MP method for functioning ULA. The SNR in this case is taken as 24 dB, the array size is 100 elements. It can be observed form Table 4, the proposed technique performs better than the TLS-MP method. The improvement is due to iteratively using the SVD. However, the improvement is obtained at the cost of slightly increased complexity.

Table 4. Estimated DOA in degrees using MP method and Modified MP method and all the 100 elements are functioning

| Assumed DOAs in degrees | Algorithm | | | |
|---|---|---|---|---|
| | MP method | | Modified MP method | |
| $\theta$ | $\hat{\theta}$ | RMSE | $\hat{\theta}$ | RMSE |
| 0 | 0.0067 | | 0.0004 | |
| 5 | 5.0024 | | 5.0001 | |
| 10 | 9.9990 | 0.0022 | 9.9948 | 0.0019 |
| 15 | 15.0019 | | 15.0014 | |
| 20 | 20.0060 | | 20.0045 | |
| 30 | 29.9967 | | 29.9984 | |

In our next example, we compare TLS-MP method and the proposed MP technique. It is assumed that, all the elements are functioning properly in case of TLS-MP method, however, 5 elements at random locations are assumed to have failed for the case of modified MP technique. We also assume only one snapshot to estimate the DOA and 20 iterations are considered for the simulation. From RMSE plot shown in Figure 3, we observe that the performance of the proposed technique with 5 faulty elements is comparable with the actual MP method, where all the elements are functioning. The results are also tabulated in Table 5, for the SNR of 24 dB.

In our third example, we consider the performance of the algorithm for varying number of working elements. The numbers of elements that are working in each experiment are 95, 90, 85, 80 and 70 elements respectively and the results are plotted in RMSE plot, shown in Figure 4. It is observed that the algorithm is consistent in its performance until 80 elements are functioning. Furthermore, when only 70 elements are functioning the performance degrades. The time taken to perform the matrix completion is observed to be 0.48 seconds and the error in reconstruction is $10^{-2}$, using 50 iterations for the matrix completion

Table 5. Estimated DOA in degrees using MP method and Modified MP method. For modified MP method 5 elements are malfunctioning

| Assumed DOAs in degrees | Algorithm | | | |
|---|---|---|---|---|
| | MP method | | Modified MP method | |
| $\theta$ | $\hat{\theta}$ | RMSE | $\hat{\theta}$ | RMSE |
| 0 | 0.0067 | | 0.0027 | |
| 5 | 5.0024 | | 5.0014 | |
| 10 | 9.9990 | 0.0022 | 10.0014 | 0.0023 |
| 15 | 15.0019 | | 14.9976 | |
| 20 | 20.0060 | | 19.9965 | |
| 30 | 29.9967 | | 29.9991 | |

## 4.1 Computational complexity

The computational complexity of the algorithms is shown in Table 6. It is observed that even though the proposed technique has more complexity, the tradeoff is that on the cost of

Signal & Image Processing : An International Journal (SIPIJ) Vol.2, No.3, September 2011

computations, we can save the use of redundant hardware when there is a failure of elements. The increase in the complexity when compared with the TLS-MP method is only due to the iterative application of SVD in MC procedure. The computational complexity in SVT is just the number of iterations multiplied by the complexity of finding the largest singular vectors.

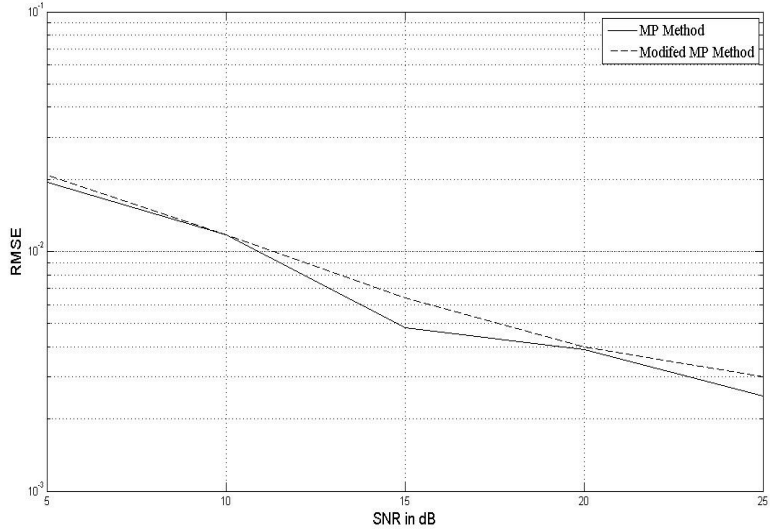

Figure 3: RMSE plot for MP method and the Modified MP method. Modified MP method is considered for faulty array, where 5 elements are failed to work.

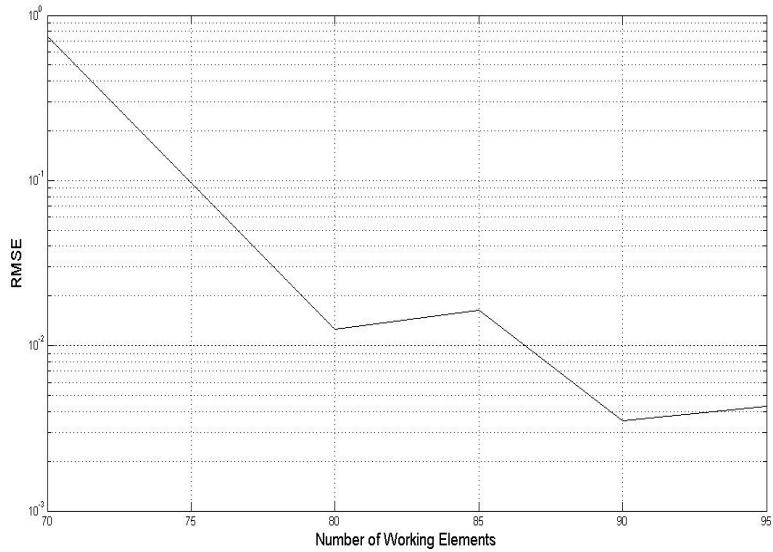

Figure 4: RMSE plot for Modified MP method for varying number of functioning elements.

Table 6. Computational Complexity

| MP Method | | Proposed Modified MP Method | |
|---|---|---|---|
| SVD | $O(4(N-L)^2(L+1) + 8(N-L)(L+1)^2 + 9(L+1)^3)$ | SVT | # of Iteration $\times\, O\left(4N\sqrt{N} + 8(\sqrt{N})^3\right)$ |
| EVD | $O((L+1)^3)$ | TLS-MP | $(4(N-L)^2(L+1) + 8(N-L)(L+1)^2 + 10(L+1)^3)$ |
| Total | $O(4(N-L)^2(L+1) + 8(N-L)(L+1)^2 + 10(L+1)^3)$ | Total | $(4(N-L)^2(L+1) + 8(N-L)(L+1)^2 + 10(L+1)^3) + 4N\sqrt{N} + 8(\sqrt{N})^3$ |

## 5. CONCLUSION

Failure of a few elements in the antenna array is not tolerated in DOA estimation applications. This paper discusses a technique to estimate the DOAs, when a few elements in the ULA fail to work. The technique uses a preprocessing scheme namely, singular value thresholding based matrix completion procedure for recovering the missing data from the failed elements. Later the MP method is employed to estimate the DOA from the recovered complete data matrix. We have also proposed a very simple technique to know the location of the elements that are working, which is required for matrix completion procedure. We evaluated the performance of the proposed modified MP method under various noisy conditions and element failure scenarios. Before considering the performance evaluation for the faulty ULA, from the simulation results, the algorithm is observed to perform better than the conventional TLS-MP method. When the elements are failed the proposed algorithms continue to estimate the DOAs until certain number of elements malfunctioning. When a large number of elements have failed the performance of the algorithm degrades. The advantage of the proposed modified MP method is usage of redundant hardware is avoided with increase in the complexity of the algorithm due to the repetitive usage of SVD. Further work has to be done to build more robust DOA estimation algorithms to tolerate the element failure, as in many applications, all the elements in the underlying ULA can rarely be expected to function properly. Direct data domain methods like Matrix Pencil method and matrix completion from the convex optimization theory offer elegant possibilities for developing robust DOA estimation algorithms.

## Acknowledgments

The useful comments made by the reviewers and Mr. Ravindra. S of Jawaharlal Nehru national college of Engineering, Shivamoga, Karnatka, India are gratefully acknowledged.

Signal & Image Processing : An International Journal (SIPIJ) Vol.2, No.3, September 2011[4]   A.P. Dempster, N.M. Laird and D.B. Rubin, (1977) "Maximum Likelihood from incomplete data via the EM Algorithm", Royal Statistical Society, Series B, Vol. 39, pp.1–38.

[5]   R. O. Schmidt, (1986) "Multiple Emitter Location and Signal Parameter Estimation", IEEE Trans. Antennas and Propagation., Vol. AP 34, pp 276-280.

[6]   R. Roy and T. Kailath, (1989) "ESPRIT - Estimation of Signal Parameters via Rotational Invariance Techniques", IEEE Trans. ASSP, Vol. 37, pp- 984–995, 1989.

[7]   Evans J.E., Johnson J.R. and Sun D.E., (1982) "Application of Advanced signal processing Techniques to Angle of Arrival Estimation", in *ATC Navigation on Surveillance system, Tech. Report 582*, MIT Lincoln Lab, Lexington, MA.

[8]   Michael Rubsamen and Alex B. Gershman, (2009) "Direction–of-Arrival Estimation for Nonuniform Sensor Arrays: From Manifold Separation to Fourier Domain MUSIC Methods", *IEEE Trans. Signal Processing*, Vol. 57, No. 2, pp. 588-599.

[9]   T. Li and A. Nehorai, (2011) "Maximum Likelihood Direction finding in Spatially Colored Noise Fields using Sparse Sensor Arrays", *IEEE trans. Signal Processing*, Vol. 59, No. 3, pp 1048-1062.

[10]  Wan B., Wan M Yakoob. M. Mohd and Sapri M. (2011), "Ubiquitous Positioning: A taxomony for location determination on mobile navigation System", *Signal and Image Processing: An International Journal,* Vol. 2,No. 1, pp-24-34.

[11]  Hua Y and Tapan K Sarkar, (1990) "Matrix Pencil Method for Estimating Parameters of Exponentially Damped/Undamped Sinusoids in Noise", *IEEE Trans. ASSP*, Vol. 38, No. 5, pp. 814-824.

[12]  Tapan K. Sarkar and Odilon Pereira, (1995) "Using the Matrix Pencil Method to Estimate the Parameters of a Sum of Complex Exponential", *IEEE Magazine, Antenna and Propagation*, Vol. 37, No. 1, pp. 48-55.

[13]  Tapan K. Sarkar, Nagaraja S. and Wicks M.C., (1998) "A Deterministic Direct Data Domain Approach to Signal Estimation Utilizing Nonuniform and Uniform 2D Array", *Digital Signal Processing*, Vol. 8, No. 2, pp. 114–125.

[14]  E. G. Larsson and P. Stoica, (2001) "High-Resolution Direction Finding: The Missing Data Case," IEEE Trans. Signal Processing, Vol. 49, No. 5, pp. 950–958.

[15]  S. Vigneshwaran, N. Sundararajan and P. Saratchandran, (2007) "Direction of Arrival (DOA) Estimation under Array Sensor Failures Using a Minimal Resource Allocation Neural Network", IEEE Trans. Antennas and Propagation, Vol. 55, No. 2, pp. 334 – 343.

[16]  Yerriswamy T. and S.N. Jagadeesha and Lucy J. Gudino, (2010) "Expectation Maximization - Matrix Pencil Method for Direction of Arrival Estimation", Proc. of 7$^{th}$ IEEE, IET International Symposium on CSNDSP 2010, UK , pp. 97–101.

[17]  E. J. Candes and B. Recht, (2008) "Exact Matrix Completion via Convex Optimization", *arxvi:0805.4471*.

[18]  B. Savas, D. Lindgren, (2006) "Rank reduction and volume minimization approach to state-space subspace system identification", Signal Processing, vol. 86, no. 11, pp- 3275–3285.

[19]  R. H. Keshavan, A. Montanari and S. Oh, (2009) "Matrix Completion from a few Entries", *arxiv:0901.3150.*

[20]  S.O. Al-Jazzara, D.C. McLernonb, M.A. Smadi, (2010) "SVD-based joint azimuth/elevation estimation with automatic pairing", Signal Processing, Vol. 90, no. 5, pp-1669–1675.
66

## Authors

**Yerriswamy T.** received B.E. in Electronics and Commn. Engg. from Gulbarga University, Karnataka, India in 2000. He received his M.Tech. in Network and Internet Engg. from Visvesvaraya Technological University, India in 2006. He is currently working towards PhD from Visvesvaraya Technological University, India. At present he is an Asst. Proff. at Proudhadevaraya Institute of Technology (affiliated to Visvesvaraya Technological University), Hosapete, Karnataka, India.

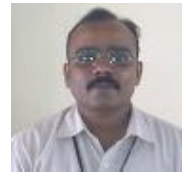

**S.N. Jagadeesha** received his B.E. in Electronics and Commn. Engg. from B.D.T College of Engg., Davanagere affiliated to Mysore University, Karnataka, India in 1979, M.E. from Indian Institute of Science, Bangalore specializing in Electrical Commn. Engg., India in 1987 and Ph.D in Electronics and Computer Engg. from University of Roorkee, Roorkee, India in 1996. He is an IEEE member. His research interest includes Array Signal Processing, Wireless Sensor Networks and Mobile Communications. He has published and presented many papers on Adaptive Array Signal Processing and Direction-of-Arrival estimation. Currently he is professor in the department of Computer Science and Engg, Jawaharlal Nehru National College of Engg. (affiliated to Visvesvaraya Technological University), Shivamogga, Karnataka, India.

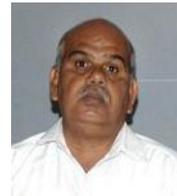